\documentclass[12pt,epsf,epsfig,psfig]{article}
\usepackage{graphicx}
\usepackage{epsfig}
\usepackage{cite}
\def\e{\epsilon}

\def\B{\boldmath}

\def\be{\begin{equation}}
\def\ee{\end{equation}}

\def\lsim{\raise0.3ex\hbox{$<$\kern-0.75em\raise-1.1ex\hbox{$\sim$}}}
\def\gsim{\raise0.3ex\hbox{$>$\kern-0.75em\raise-1.1ex\hbox{$\sim$}}}
\def\NP{{ Nucl.\ Phys.\ }}
\def\PL{{ Phys.\ Lett.\ }}
\def\PR{{ Phys.\ Rev.\ }}

\def\PRL{{ Phys.\ Rev.\ Lett.\ }}

\def\ZP{{ Z.\ Phys.\ }}
\def\EP{{ Europ.\ Phys.\ J.\ C}}

\begin{document}

11.3.11 \hfill BI-TP 2011/01

\vskip 0.5cm

\centerline{\Large \bf Trace Anomaly and Quasi-Particles}

\medskip

\centerline{\Large \bf in Finite Temperature \B$SU(N)$ Gauge Theory}

\vskip 0.8cm 

\centerline{\large \bf P.\ Castorina$^a$, D.\ E.\ Miller$^{b,c}$ and 
H.\ Satz$^c$} 

\bigskip

\centerline{a) Dipartimento di Fisica, Universit{\`a} di Catania,}

\centerline{and INFN Sezione di Catania, I-95123 Catania, Italia}

\medskip

\centerline{b) Department of Physics, Pennsylvania State University}

\centerline{Hazleton Campus, Hazleton, PA 18202 USA}

\medskip

\centerline{c) Fakult\"at f\"ur Physik, Universit\"at Bielefeld, 
D-33501 Bielefeld, Germany}

\vskip1cm

\centerline{\large \bf Abstract}

\bigskip

We consider deconfined matter in $SU(N)$ gauge theory as an ideal
gas of transversely polarized quasi-particle modes having a 
temperature-dependent mass $m(T)$. Just above the transition temperature, 
the mass is assumed to be determined by the critical behavior of the 
energy density and the screening length in the medium. At high
temperature, it becomes proportional to $T$ as the only remaining
scale. The resulting (trace anomaly based) interaction measure 
$\Delta=(\e - 3P)/T^4$ and energy density are found to agree well 
with finite temperature $SU(3)$ lattice calculations.  

\vskip1cm

\section{Introduction}

The quark-gluon plasma in the region $T_c \leq T \leq 5~\!T_c$  
presents a particularly challenging topic of investigation to
strong interaction thermodynamics. The most suitable tool
for such studies is the expectation value of the
trace of the energy-momentum tensor, $\langle \Theta_{\mu}^{\mu}\rangle
 = \e - 3~\!P$, which measures the deviation
from conformal behavior and thus identifies the interaction still
present in the medium. The aim of the present work is to study
the temperature behavior of this measure and try to identify the
underlying physics which causes it. The only {\sl ab initio}
calculations in the range of temperatures of interest here are
obtained through finite temperature lattice QCD. In particular,
pure $SU(3)$ gauge theory has been studied extensively, and through 
finite size scaling techniques the behavior is given in the continuum 
limit \cite{Boyd}. This case will therefore form the main basis of our 
study.

\medskip

\begin{figure}[h]
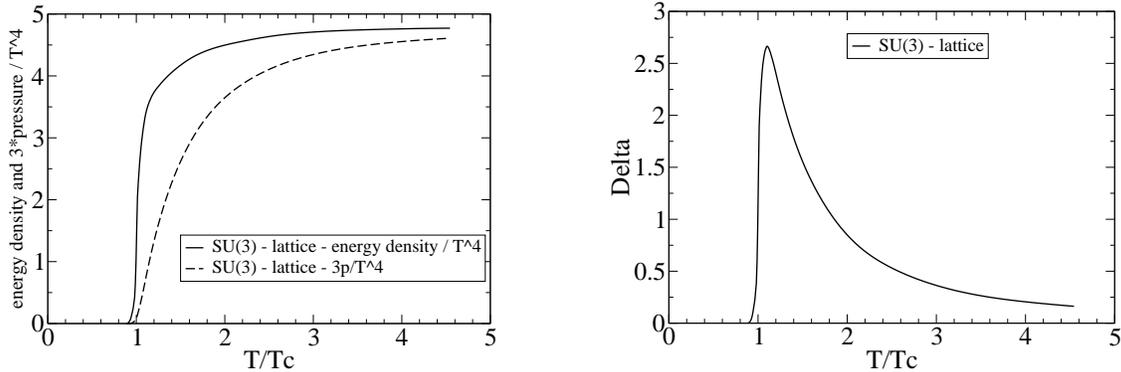

\centerline{\psfig{file=energypressuresu3L.eps,width=6.5cm}\hskip1.5cm
\psfig{file=deltasu3dataL.eps,width=6.8cm}}
\caption{Temperature dependence (left) of the energy density and the pressure, 
and (right) of the interaction measure, for $SU(3)$ gauge theory \cite{Boyd}.}
\label{EDP}
\end{figure}

In Fig.\ \ref{EDP}a we show the temperature dependence of the energy  
and the pressure, divided by $T^4$, for SU(3) gauge theory, and 
in Fig.\ \ref{EDP}b that of the dimensionless interaction measure, 
\be
\Delta(T) = {\langle \Theta_{\mu}^{\mu} \rangle \over T^4}=
{\e - 3~\!P \over T^4}.
\label{delta}
\ee 
For all quantities, the extrapolation to the continuum limit is shown
\cite{Boyd}.  
In the region around and just above $T_c$, the energy density rises
much more rapidly than the pressure, leading to the observed rapid
increase of $\Delta(T)$. Since asymptotically $\e(T)/T^4$ and 
$3~\!P(T)/T^4$ converge to their common Stefan-Boltzmann value $(8~\! 
\pi^2/15)$, there must be some
temperature $T_p$ at which the growth rates change roles, with the
pressure now increasing more rapidly. This leads to the peak observed
for $\Delta(T_p \simeq 1.05~\!T_c)$, followed by a somewhat slower decrease. 
The transition itsself is of first order \cite{Celik83,Kogut83}, as 
expected for a theory belonging to the $Z_3$ universality 
class \cite{Sve-Y82}. 

\vskip0.5cm

\medskip

\begin{figure}[h]
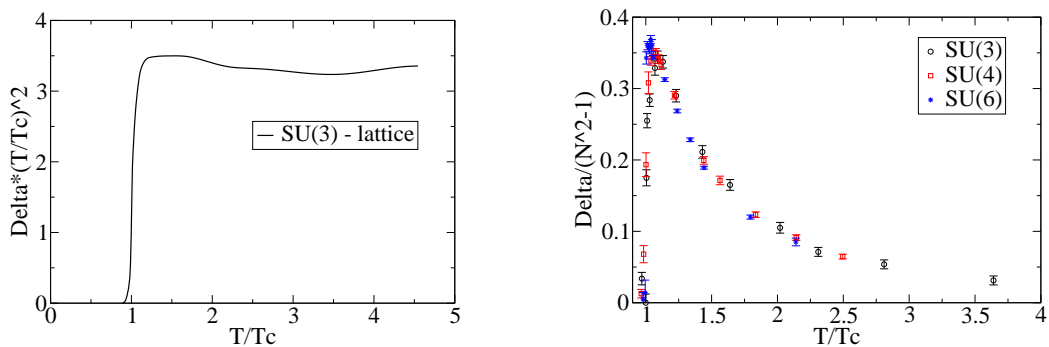

\centerline{\psfig{file=su3scaled.eps,width=6cm}\hskip1.5cm
\psfig{file=scalingcernL.eps,width=6.3cm}}
\caption{Temperature dependence (left) of $(T/T_c)^2~\Delta(T)$ for
$SU(3)$ gauge theory \cite{Boyd} and (right) of the scaled interaction measure
$\Delta(T)/(N^2-1)$ for $SU(N)$ gauge theories with $N=3,4,6$.} 
\label{T2}
\end{figure}

Two further features of the interaction have become clear more recently. As 
shown in Fig.\ \ref{T2} (left), the decrease of $\Delta$ in the region under
consideration here is in good approximation given by $T^{-2}$, so that
$T^2 \Delta(T)$ becomes approximately constant very soon above $T_c$, and
up to about $5~\!T_c$ \cite{T-2}. Moreover,
it is seen that $\Delta(T)$ in different $SU(N)$ theories scales very 
well with the number of gluonic degrees of freedom \cite{datta-gupta};
in other words, $\Delta(T)/(N^2 -1)$ becomes a universal curve, as seen 
in Figure \ref{T2} (right).

\vskip0.5cm

\medskip

In the following, we will first check to what extent any of the observed
behavior can be accounted for by conventional or modified perturbation
theory. Next we turn to the non-perturbative approach obtained through
the relation of the interaction measure with the gluon condensate and
the corresponding bag pressure. Finally, in the main part of our work,
we shall then show that a quasi-particle approach, based on massive
gluonic modes, can indeed provide an excellent description of all the
features observed in numerical finite temperature lattice studies
of $SU(3)$ gauge theory .

\section{Weak Coupling Approaches}

In perturbation theory, the interaction measure for pure $SU(N)$ gauge 
theory is to leading order given by \cite{Boyd,Kapusta}
\be
\Delta_{\rm pert}={(N^2-1)\over 28 8}~\!{11 \over 12 \pi^2}~\! N^2 g^4(T),
\label{Dpert}
\ee
with
\be
N~\!g^2(T) = {24~\! \pi^2 \over 11 ~\!\ln(T/\Lambda_T)}.
\label{gpert}
\ee
The perturbative interaction measure thus does show the observed 
scaling in $N^2-1$ just mentioned, assuming $N~\!g^2$ is kept constant.
In eq.\ (\ref{gpert}), $\Lambda_T$ defines the lattice scale, which in 
the mentioned
$SU(3)$ lattice studies \cite{Boyd} was found to be determined by 
$T_c/\Lambda_T = 7.16 \pm 0.25$. In this case, we thus obtain
\be
\Delta_{\rm pert}= {11 \over 48 \pi^2}~\!g^4=
{4 \pi^2 \over 33}~ {1 \over \{\ln [7.16(T/T_c)]\}^2}
\simeq {1.2 \over \{\ln [7.16(T/T_c)]\}^2}. 
\label{gpert3}
\ee
At $T/T_c = 3$, we have $\Delta_{pert} \simeq 0.13$, which is
still about a factor 3 below the (continuum extrapolated) lattice result 
$\Delta_{lat} \simeq 0.4$. Hence at this temperature, leading order
perturbation theory cannot yet reproduce the plasma interaction. 
Nevertheless, we have here $\alpha_s=g^2/4\pi \simeq 0.19$ for the strong 
coupling $\alpha_s$, so that in principle perturbation theory seems to be 
applicable, and we could expect that at somewhat higher temperatures,
above $T/T_c \simeq 5-10$, the perturbative form might account for 
the lattice result. 

\medskip

\begin{figure}[h]
\centerline{\psfig{file=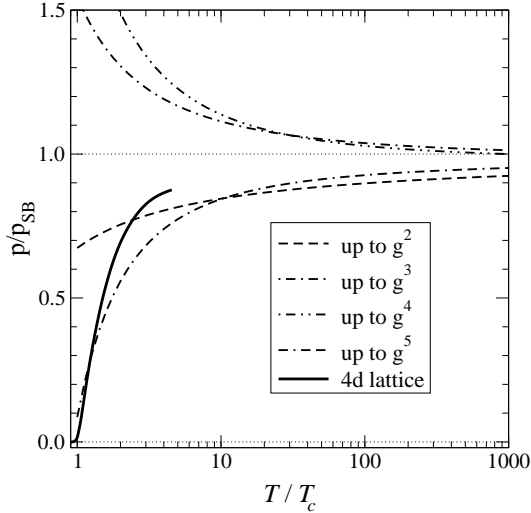,width=7cm}}
\caption{Perturbative expansions of the pressure in $SU(3)$ gauge
theory \cite{Arnold94,Laine}, compared to the finite temperature lattice 
results \cite{Boyd}.}
\label{mikko}
\end{figure}

The evaluation of higher order perturbative terms has, however, shown that
this is not the case. Infrared divergences in finite temperature field 
theory limit calculations to a finite order in the coupling $g$ \cite{Linde}; 
for the pressure, the highest perturbatively calculable order is $g^5$, 
and calculations have now been extended to this order \cite{Arnold94}. In 
Fig.\ \ref{mikko}, we show the result of expansions in different order 
$g^n$ for the pressure in $SU(3)$ gauge theory, normalized to 
the Stefan-Boltzmann limit\cite{Laine}. It is seen that in the temperature 
region of interest here, $T \leq 10~T_c$, the different orders lead to 
strong fluctuations; the final form, up to and including $O(g^5)$, still 
considerably undershoots the lattice results. 

\medskip

\medskip

Moreover, for an understanding of the interaction effects, a comparison 
of lattice and perturbation theory results for the {\sl pressure} is in fact
quite misleading, since the major part of the pressure is given by
the ideal gas component. To concentrate on just the interaction effects,
we return to the interaction measure $\Delta(T)$, and here perturbation 
theory breaks down completely. The next-to-leading order (NLO) form for 
$SU(3)$ gauge theory,
\be
\Delta_{pert}= \left({11 \over 8 \pi^2}\right)
\left[{1\over 6} g^4 - {1\over \pi} g^5\right]
\label{twoloop}
\ee
becomes positive only for $g^2 \gsim 0.27$, which with the two-loop
form of the coupling,
\be
g^{-2}={11\over 8\pi^2}~\ln(T/\Lambda_T) + {51\over 88 \pi^2}~
\ln[2~\ln(T/\Lambda_T)],
\label{twoloop-g}
\ee
requires inconceivably high temperatures, above $10^6 ~T_c$. We conclude 
that the interaction of the plasma in the region of interest here, up
to some 10 $T_c$, must definitely require some non-perturbative features.

\medskip

\begin{figure}[h]
\centerline{\psfig{file=pertg8.eps,width=7cm,height=5cm}\hskip1cm
\psfig{file=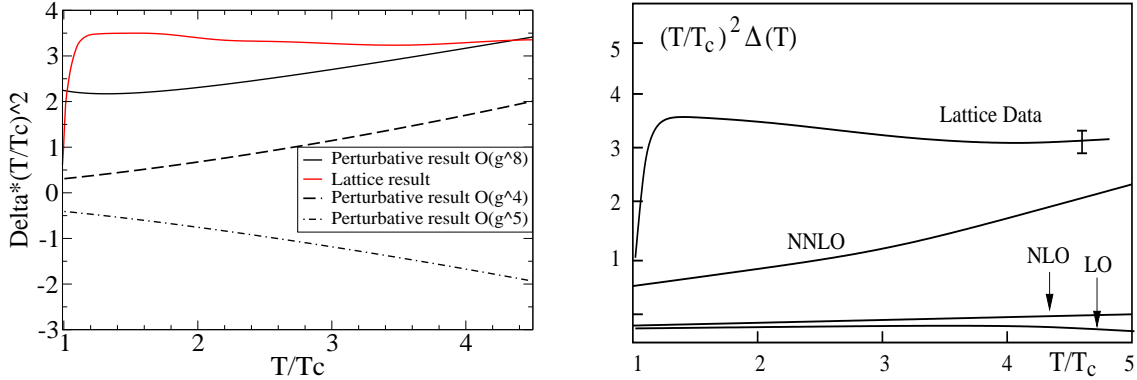,width=7cm,height=5cm}}
\caption{$(T/T_c)^2 \Delta(T)$ as predicted in systematic effective
field theory (left, \cite{Laine}) and in HTL resummed perturbation theory 
(right, \cite{HTL}), compared to the continuum extrapolation
of lattice studies \cite{Boyd}.} 
\label{Trace}
\end{figure}

This situation has triggered numerous efforts to modify the perturbative
approach to include such features. In one approach \cite{Laine}, the $O(g^6)$ 
term in the pressure is evaluated by a non-perturbative scale determination,
using lattice results for magnetic screening. This leads to a systematic
effective field theory, for which in principle all orders can be calculated.
Another possibility is given by including sums over certain graph classes, 
thus effectively shifting the point about which the perturbation expansion 
is performed \cite{Patkos}. In particular, this is studied for the terms 
dominating at
high temperature (hard thermal loops, HTL) \cite{HTL,resum} and leads to 
an improved convergence of the perturbation series of the pressure. 
Both approaches have in common 
\begin{itemize}
\item{a rather good description of the pressure for temperatures
above 3 -5 $T_c$, but}
\item{the range below about 3 $T_c$ is still not well accounted for.}
\item{Moreover, the strong order-by-order fluctuations for the interaction
measure cause some doubt that the last order considered is really close to 
a ``final'' result.}
\end{itemize}
To illustrate this, we show in Fig.\ \ref{Trace} (left) the behavior obtained
with the help of a partially non-perturbative $O(g^6)$ term\cite{Laine},
and in Fig.\ \ref{Trace} (right) corresponding results from modified HTL 
calculations \cite{HTL}, in both cases compared to the form obtained in 
$SU(3)$ lattice QCD. The latter show for $\e - 3P$ a decrease as $1/T^2$, 
so that $T^2 \Delta(T)$ becomes approximately constant above $T_c$. We see
in Fig.\ \ref{Trace} (right) that in leading (LO) and next-to-leading order 
(NLO) the breakdown of perturbation theory persists also in a HTL approach, 
and even the inclusion of a partially non-perturbative NNLO contribution
cannot reproduce the lattice result, neither in size nor in functional
form. Such a conclusion had been reached before, see e.g.\ \cite{L-Z}.
Recent studies \cite{Strick2} have shown that in the case of full QCD 
with light quark flavors, the discrepancy between lattice data and
weak-coupling results is reduced, with quite good agreement down to 
about 2 - 3 $T_c$; however, neither the approximate $T^{-2}$ behavior 
of $\Delta(T)$ in the range from $T_c$ to about 5 $T_c$, nor the sudden 
drop in the critical region can be thus obtained.

\bigskip
 
\medskip

In general, the breakdown observed in any perturbative treatment as we 
enter the transition region is of course not surprising. Critical or 
even pseudo-critical behavior with an increasing correlation range is 
simply not a perturbative phenomenon. We therefore have to find a
non-perturbative approach to address the behavior of the plasma in 
this region.

\medskip

\section{Bag pressure and Gluon Condensate}

One of the earliest attempts to account for the essential non-perturbative
features is provided by the bag model \cite{bag,Baacke}. Here one implements 
confinement in an ideal gas picture by introducing a bag pressure, measuring 
the ``level difference'' between the physical vacuum and the ground state
in the colored world of QCD. For the corresponding 
thermodynamics,
this means that to the ideal gas partition function, $Z_0(T,V)$, a bag term 
is added,
\be
T \ln Z_B(T,V) = T \ln Z_0(T,V) - BV
\label{bag}
\ee
which in principle can be determined from a bag model description of 
hadron spectroscopy. The bag pressure simulates a form of 
interaction \cite{AH}, 
as best seen by the resulting interaction measure,
\be
\Delta(T) = {4~\!B \over T^4}.
\label{deltabag}
\ee
We want to check here to what extent this is a viable description 
of the QGP interaction in the region above $T_c$, assuming $B$ to
remain temperature-independent. 

\medskip

The thermal expectation value of the trace of the energy-momentum tensor, 
$\langle \Theta_{\mu}^{\mu}\rangle = \e - 3 P$, is related to the gluon 
condensate, i.e., to the expectation value of gluon term in the QCD 
Lagrangian \cite{Leut},
\be
G^2 \equiv {\beta(g) \over2~\! g^3}  G_{\mu \nu}^a G^{\mu \nu}_a
= {11 N_c \over 96 \pi^2} G_{\mu \nu}^a G^{\mu \nu}_a,
%= - {11 (g^2 N_c) ỏver 96 \pi^2} F_{\mu \nu}^a F^{\mu \nu}_a,
\label{glu1}
\ee
where $G_{\mu \nu}^a =g F_{\mu \nu}^a$ is given by the gluon field of 
color $a$ in the QCD Lagrangian. The last line of eq.\ (\ref{glu1})
is obtained using the leading order perturbative beta function,
\be
\beta (g) = {11~\!N_c \over 48 \pi^2} ~\!g^3 + O(g^5).
\label{beta}
\ee
The value of $\langle G^2 \rangle = G^2_0$ at $T=0$ has been estimated
numerically, with $G_0^2=0.012 \pm 0.006$ GeV$^4$ as ``canonical'' value
\cite{SVZ}. In both analytical \cite{Leut} and in lattice studies
\cite{Boyd}, $\e - 3P$ is normalized to zero at $T=0$, so that
\be
\langle \Theta_{\mu}^{\mu}\rangle= \e - 3P = G_0^2 - G_T^2,
\label{glu2}
\ee
where $G_T^2$ is the temperature-dependent gluon condensate. In the
temperature range below $T_c$, we expect $G_T^2=G_0^2$, so that
$\e - 3P=0$. If the gluon condensate melts above $T_c$, the
interaction measure becomes
\be
{\e - 3P \over T^4} = {G_0^2 \over T^4},
\label{glu3} 
\ee
so that $B=G_0^2/4$ relates bag pressure and gluon condensate. The
value for the latter given above leads to a bag pressure $B^{1/4} 
\simeq 230 \pm 30$ MeV, which is in reasonable agreement with that 
obtained from hadron spectroscopy as given by the bag model.  

\medskip

The color summation in eq.\ (\ref{glu1}) runs over the $N^2 - 1$
gluonic color degrees of freedom, so that we can write
\be
G^2 = 
{11~\!N~\! g^2 \over 96 \pi^2}\langle F_{\mu \nu}^a F_a^{\mu \nu} \rangle
= {11~\!N~\! g^2 \over 96 \pi^2}(N^2 -1) \langle{\bar F}_{\mu \nu} 
{\bar F}^{\mu \nu} \rangle,
\label{glu5}
\ee
where $\langle{\bar F}_{\mu \nu} {\bar F}^{\mu \nu} \rangle$ denotes
the gluon field contribution per color degree of freedom.
The scaling of the interaction measure in $N^2-1$ observed for different
$SU(N)$ theories is thus also in accord with the bag model dependence,
keeping $g^2N$ constant. 

\medskip

\begin{figure}[h]
\vskip0.5cm
\centerline{\psfig{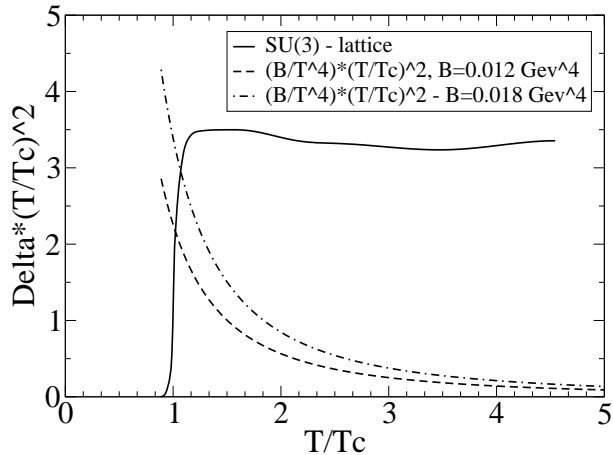}}
\caption{The temperature variation of $\Delta(T)(T/T_c)^2$ obtained
from the bag pressure, compared to the corresponding lattice data
\cite{Boyd}.}
\label{bag-delta}
\end{figure}

If one assumes that $G_T^2=0$ for $T\geq T_c$\footnote{It is known 
that such an assumption is in general not tenable \cite{DEM}; 
a description in terms of a temperature-independent bag constant
must therefore fail eventually. We want to determine here if it
makes sense anywhere.} the interaction measure
becomes
\be
\Delta(T) = {G_0^2 \over T^4} = 
{G^2_0 \over T^4_c}\left({T_c\over T}\right)^4 \simeq {2.3 \over (T/T_c)^4},
\label{glu4}
\ee
using $T_c \simeq 0.27$ GeV for the $SU(3)$ deconfinement temperature.
The lattice data are found to decrease much slower and are, as mentioned,
in accord with a $1/T^2$ dependence. We therefore compare in
Fig.\ \ref{bag-delta} the results for $T^2 \Delta(T)$ given by
the lattice and the bag model forms. The bag model naturally cannot 
account for the structure immediately around  $T_c$ (the rise to the 
peak of $\Delta(T)$), but it also fails in the temperature region 
above $T_c$. Combining the bag model with some form of weak-coupling
appproach can somewhat improve the latter, but it can never reproduce
the behavior in the critical region. This remains true also in various
other, conceptually interesting attempts to modify the power of the 
$T$-dependence of $\Delta(T)$ \cite{Zwanziger,Rob,Megias}.

\medskip

\medskip

\section{The Quasi-Particle Approach}

There thus remains the task to find a 
non-perturbative approach which takes into account 
the critical features arising in the temperature region in the range 
above $T_c$, as they were obtained in lattice studies. We stay in 
pure $SU(3)$ gauge theory, where, as mentioned, 
the extrapolation to the continuum
limit is known \cite{Boyd}. The basis for our considerations here
is the study of an ideal gas of constituents 
(``quasi-particles'') having dynamically or thermally generated masses 
\cite{Golo93,Peshier,Brau}. The behaviour of an ideal gas of such massive 
gluon modes provides automatically the observed $N_c^2$ scaling and 
also leads to other features in accord with the functional behaviour 
found in $SU(N)$ gauge theories. 

\medskip

Interpretations of lattice QCD results in terms of a quasi-particle
picture have been given in many versions \cite{Golo93,Peshier,Brau,LH,Gia}.
In our approach, as in \cite{Golo93,Brau}, we shall include all
interaction effects in a dynamically generated mass $m(T)$; most 
other studies maintain in addition a temperature-dependent bag constant.
Instead, we want to relate the behavior near $T_c$ to the critical
behavior of the correlation length, causing the effective 
mass to increase as $T \to T_c$ from above \cite{Golo93}.   

\medskip

The partition function of an ideal gas of constituents of mass $m(T)$ is 
in the Boltzmann limit for $SU(N)$ gauge theory given by
\be
\ln Z(T) = 2{(N^2-1) V\over 2 \pi^2}
\int_0^{\infty} dp~p^2 ~\!\exp(-{1\over T} \sqrt{p^2 + m^2}) 
= 2{(N^2-1) V Tm^2 \over 2 \pi^2}K_2(m/T),
\label{M-1}
\ee 
where $K_i(x)$ denotes the Hankel function of imaginary argument.
The resulting pressure becomes
\begin{eqnarray}
P(T) =T \left({\partial \ln Z \over \partial V}\right)_T
\hspace*{4cm}
\nonumber \\
= 2{(N^2-1) T \over 2 \pi^2}
\int_0^{\infty} dp~p^2 ~\!\exp(-{1\over T} \sqrt{p^2 + m^2}) 
= 2{(N^2-1) T^2m^2 \over 2 \pi^2}K_2(m/T)
\label{pressure}
\end{eqnarray}
while the energy density is found to be
\begin{eqnarray}
\e(T) = {T^2 \over V}  \left( {\partial \ln Z(T) \over \partial T} 
\right)_V \hspace*{5cm} \nonumber\\
= 2{(N^2-1) \over 2 \pi^2}
\int_0^{\infty} dp~p^2 ~\!\exp(-{1\over T} \sqrt{p^2 + m^2})
\left\{ \sqrt{p^2 + m^2} - T {m \over  \sqrt{p^2 + m^2}}
\left({d m \over dT}\right) \right\}               
\nonumber\\
= 2{(N^2-1) m^2 T^2 \over 2 \pi^2}
\left\{3 K_2(m/T) + \left[{m \over T} - \left({dm \over dT}\right) \right]
K_1(m/T) \right\}\hspace*{2cm}
\label{endensity}
\end{eqnarray}
In these expressions, we have maintained two spin degrees of freedom
for the ``massive'' gluons; we return to this point shortly.
Both energy density and pressure thus fall below the Stefan-Boltzmann
limit, as is observed in the lattice data shown in Fig.\ \ref{EDP}. The 
resulting interaction measure is given by
\be
\Delta(T) = 2{(N^2-1) \over 2 \pi^2 T^4}
\int_0^{\infty} dp~p^2~\! \exp(-{1\over T}\sqrt{p^2 + m^2})
\left\{ \sqrt{p^2 + m^2} - 3~\!T - 
T{m \over \sqrt{p^2+m^2}}\left({dm \over dT}\right)
\right\}
$$
$$
= 2{(N^2-1) m^2 \over 2 \pi^2 T^2}
\left[{m \over T} - \left( {dm \over dT}\right) \right]
K_1(m/T) 
\label{delta-m}
\ee
If $m$ is $N $-independent, the scaling in $N^2-1$ is evident. 
Moreover, if the effective mass $m$ is linear in $T$, as in any 
conformal theory, $\Delta(T)$ vanishes. Given a running coupling, with
$m^2 = N~\!g^2(T) T^2/3$, we get
\be
\Delta(T) =2{(N^2-1) m^2 \over 12 \pi^2 T^2} T\left( {d(g^2N) \over dT}\right).
\label{delta-run}
\ee

\medskip

We note, however, that such a ``naive'' quasi-particle description with 
finite masses seems to encounter a conceptual problem. Physical 
constituents of non-vanishing mass should have three, rather than two 
spin degrees of freedom, since a longitudinal polarization is excluded 
only for massless particles. The resulting changes in all thermodynamic 
quantities -- e.g., the increase of the ideal gas energy density $\e/T^4$ 
 from $8 \pi^2/ 15$ to $12 \pi^2 /15$ -- are 
definitely in disagreement with the observed high temperature lattice results.
A simple shift to massive gluons thus cannot satisfactorily explain the 
interactions apparently still present in the high temperature gluon gas. 
More generally, a gauge invariant theory does not allow massive physical 
gluons; the mechanism leading to effective thermal masses must thus be 
more subtle. The mentioned modified HTL perturbation theory approach,
in which each order already includes some aspects of gluon dressing, 
not only leads to a rather rapid convergence of the expansion; here  
the contribution of longitudinal gluons vanishes in the limit $g \to 0$, 
so that one also obtains the right number of degrees of freedom 
for the Stefan-Boltzmann form \cite{resum}. Moreover, it has recently
been argued \cite{Mathieu} that masssive gluons should in fact be
transversely polarized, since two massless gluons cannot combine to
form a longitudinally polarized massive gluon \cite{Yang}. It thus seems
justified to use the thermal mass scenario outlined above to address the 
temperature behaviour of the quark-gluon plasma. 

\medskip

The form of the effective mass entering in a quasi-particle approach
description has been an enigma for quite some time. At sufficiently high 
temperature, $T$ remains as the only scale, so that there we expect 
$m \sim T$. From perturbation theory one obtains in leading order for 
$SU(N)$ gauge theory a thermal screening mass $m^2 \sim N~\!g^2(T)~\!T/3$, 
but in view of the above mentioned difficulties, it seems best to leave the 
proportionality open. As we approach the critical point, perturbation theory 
in whatever form ceases to be applicable. We now have a medium of strongly
interacting gluons, and the range of the forces between them becomes larger 
and larger as we approach the critical point. This range is governed by the 
correlation length, or in other words, by the distance up to which a
given color charge can ``see'' other color charges. This distance
is the QCD counterpart of the Debye screening radius in QED; we write 
it as $r_D(T)=1/\mu(T)$, where $\mu(T)$ denotes the corresponding screening 
mass. It corresponds to the shift from $1/k^2$ to $1/(k^2 + \mu^2)$ 
experienced by the gluon propagator due to the presence of the medium.

\medskip

\begin{figure}[h]
\centerline{~~~~\psfig{file=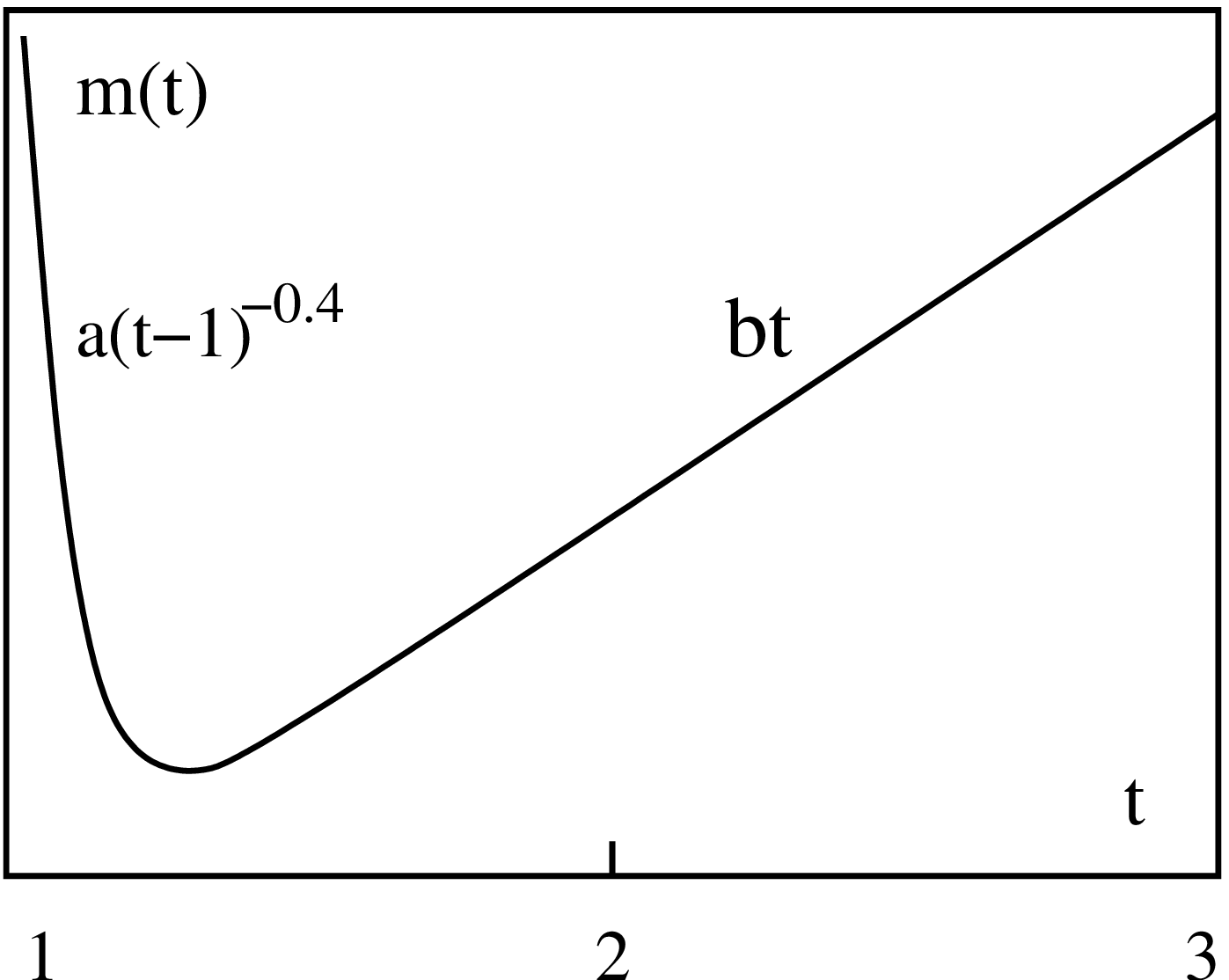,width=5.5cm,height=4cm} \hskip1.5cm
\psfig{file=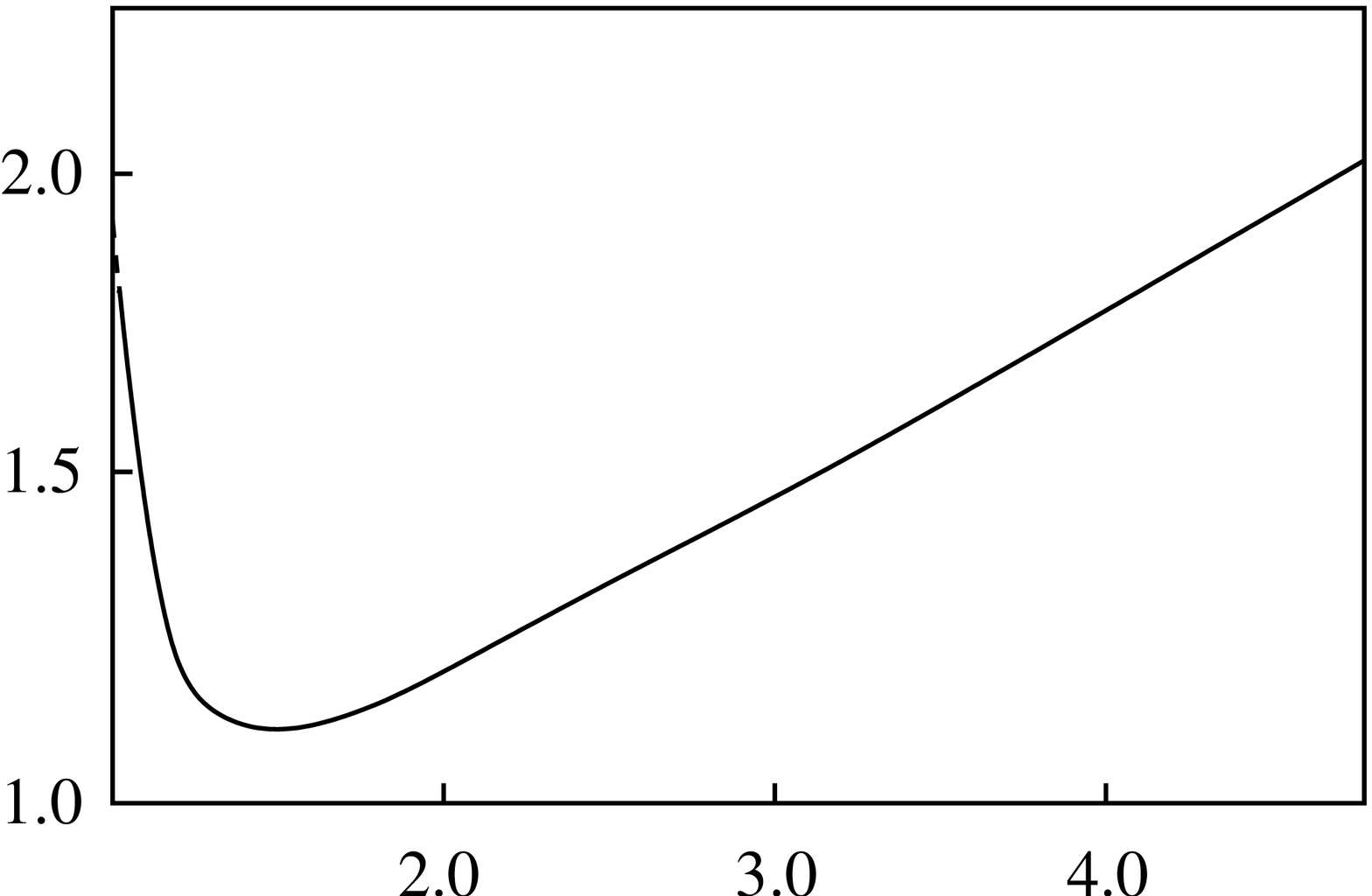,width=6cm,height=4cm}}
\caption{Left: Expected behavior of the effective quasi-particle mass $m(t)$
in (left) $SU(2)$ gauge theory, and (right) as obtained from a fit of
lattice data in $SU(3)$ gauge theory.} 
\label{critmass}
\end{figure}

The perhaps simplest view thus is to consider the mass of the quasi-gluon
in the strongly coupled region to be the energy contained in a volume 
$V_{\rm cor}$ of the size defined by the correlation range,
\be
m_{crit}(T) \sim \e(T) V_{\rm cor}(T).
\label{critmass3}
\ee
In the case of a continuous transition, the critical part of the
energy density becomes
\be
\e_{\rm crit} \sim (t-1)^{1-\alpha},
 \label{en1}
\ee
where $t=T/T_c$  and $\alpha$ is the critical exponent for
the specific heat. The correlation volume (for three space dimensions)
can be written as
\be
V_{\rm cor} = 4\pi \int dr r^2 ~\Gamma(r,T),
\label{cor1}
\ee
where
\be
\Gamma(r,T) \sim {\exp-\{r/\xi(T)\} \over r^{1-\eta}},
~~\xi(T) \sim (t-1)^{-\nu}
\label{cor2}
\ee
specifies the correlation function $\Gamma$ in terms of critical exponents
$\nu$ for the correlation length $\xi(T)$ and $\eta$ as anomalous dimension 
exponent. Combining these expressions and making use of the exponent equality
relating $\alpha$ and $\nu$, we obtain
\be
m_{crit}(T) \sim (t-1)^{1-\alpha - 2\nu - \eta} = (t-1)^{-(1+\eta-\nu)};
\label{critmass4}
\ee
for a continuous transition in three space dimensions. For $SU(2)$ gauge
theory, the critical exponents are given by the corresponding exponents 
of the 3d Ising model \cite{Sve-Y82}, $\nu \simeq 0.63$ and 
$\eta \simeq 0.04$, suggesting
\be
m_{crit}(T) \sim (t-1)^{-0.41}.
\label{critmass5}
\ee
This form is correct only in very near the critical point $t=1$; for
large temperatures, $\xi(t) \sim t$, so that the overall form expected
for the mass of the quasi-gluon becomes
\be
m(t) \simeq a(t-1)^{-0.41} + b~\!t,
\label{mass}
\ee
where $a$ and $b$ are constants. The resulting behavior is illustrated 
in Fig.\ \ref{critmass} (left). It would certainly be of interest to 
check this form directly through calculations in $SU(2)$ 
gauge theory; unfortunately, there does not seem to exist any lattice
study providing an extrapolation to the continuum, thus eliminating finite 
lattice size effects. Older studies of $\e(T)$ and $P(T)$ in terms of
a gluon mass $m(T)$ \cite{Golo93} did in fact lead to the form shown in 
Fig.\ \ref{mass}.

\medskip

For $SU(3)$, the transition is of first order \cite{Celik83,Kogut83}, 
so that all quantitites
remain finite at $T_c$ and an equivalent form cannot be given.  
Nevertheless, in all cases we have a strong increase of both $\e(T)$ 
and $\Delta(T)$ in some range above $T_c$, and so we shall maintain the 
functional dependence (\ref{critmass4}/\ref{critmass5}) with an open 
exponent $c$. The resulting quasi-particle mass is thus expected to 
have the form
\be
m(T) = {a \over (t-1)^c} + bt ,
\label{critmass6}
\ee  
with constants $a,~b,~c$. 

\medskip

\medskip

\begin{figure}[h]
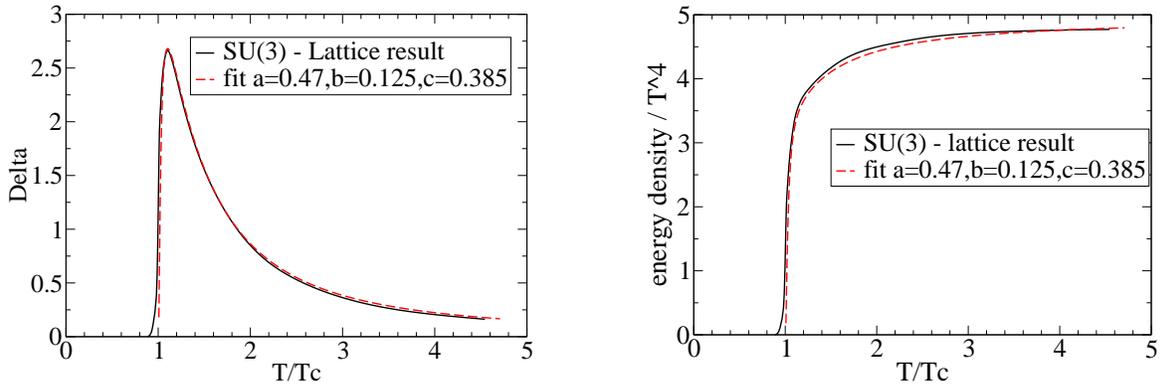

\vskip0.5cm
\centerline{\psfig{file=fitdelta.eps,width=7cm}\hskip1.5cm
\psfig{file=fitenergysu3L.eps,width=6.8cm}}
\caption{Interaction measure (left) and energy density (right)
for $SU(3)$ gauge theory, compared to a quasi-particle description;
the fitted constants $a$ and $b$ are given in GeV.}
\label{fits}
\end{figure}

Using this mass, we now determine the parameters $a,b,c$ by calculating 
the energy density from eq.\ (17) and $\Delta(T)$ from eq.\ 
(\ref{delta-run}). The resulting mass and the corresponding parameters
are shown in Fig.\ \ref{critmass} (right).
The fits to energy density and interaction measure are given
in Fig.\ \ref{fits} and are seen to reproduce both
quantities very well. We can thus conclude that the gluon plasma
in $SU(3)$ gauge theory in the temperature region above $T_c$ indeed
behaves like a medium of quasi-particles with masses generated through
non-perturbative thermal effects.

\bigskip

\subsection{The Speed of Sound in the QGP}

In a hadronic resonance gas, one finds \cite{CCMS} that the
speed of sound drops to zero at the critical point defined by
the limit of hadronic matter. It does so because any further
energy increase goes into making more massive resonances, not
into momentum and pressure. On the QGP side, in the quasi-particle
just discussed, the behavior is very similar. As we lower the
QGP temperature towards the confinement point, the increase of 
the quasi-particle mass has the same effect. In other words,
a temperature increase above $T_c$ lowers the mass and thus
provides more momentum and pressure, causing an increase in 
the speed of sound.

\medskip

\begin{figure}[h]
\vskip0.5cm
\centerline{\psfig{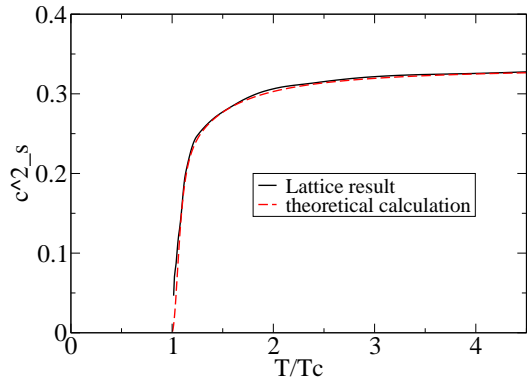}}
\caption{The speed of sound in the quasi-particle approach, compared
to the $SU(3)$ lattice result \cite{Boyd}.}
\label{QGPsound}
\end{figure}

The speed of sound, defined as
\be
c_s^2 = \left({\partial p \over \partial \e}\right)_V
= {s(T) \over C_V(T)},
\label{sound1}
\ee 
vanishes at $T_c$ for a continuous transition, because the specific heat
$C_V(T)$ diverges there, while the entropy density $s(T)$ remains finite.  
In the ideal gas limit, $s(T) \simeq 4~\!c_0~\!T^3$ and $C_V(T) \simeq
12~\!T^3$, so that $c_s^2 \to 1/3$. For the temperature-dependent mass 
(\ref{critmass6}), the speed of sound can be evaluated numerically,
using eqns.\ (\ref{pressure}) and (\ref{endensity}).
%using eqns.\ (\ref{pressure}) and (\ref{endensity}). 
In Fig.\ \ref{QGPsound}, 
we show the resulting behavior obtained in our quasi-particle approach, 
in comparison to the SU(3) lattice results. The two forms agree very well 
and in fact provide the behavior just indicated.

\section{Conclusions}

We have shown that the temperature behavior of the interaction measure
defined by the trace anomaly of the energy-momentum tensor, 
$\Delta(T)=(\e - 3P)/T^4$, is well described in terms of massive
gluons with only transverse degrees of freedom. The gluon mass $m(T)$
increases sharply as $T \to T_c$ from above, due to the rapid growth
of the correlation length in the critical region. On the other hand, 
with increasing
temperatures and the approach to conformal behavior, $m(T) \sim T$.
The combination of these two effects results in a minimum of $m(T)$
around 1.5 $T_c$, signalling the transition from critical to hot
gluon plasma. Even the hot plasma, however, remains strongly interacting;
weak-coupling studies do not reproduce the observed lattice behavior in
the region below about 5 $T_c$.

\vskip1cm

\section{Acknowledgement}

It is pleasure to thank J.\ Engels, O.\ Kaczmarek and Y.\ Schroeder
for helpful discussions and for providing essential data.


\vskip1cm


\begin{thebibliography}{99}

\bibitem{Boyd} G.\ Boyd et al., \NP B 469 (1996) 419.

\bibitem{Celik83} T.\ {\c C}elik et al., \PL B125 (1983) 411.

\bibitem{Kogut83} J.\ Kogut et al.,  \PRL 51 (1983) 869.

\bibitem{Sve-Y82} B.\ Svetitsky and L.\ Yaffe, \NP B 210 [FS6] (1982) 423.

\bibitem{T-2} R.\ Pisarski, Progr.\ Theoret.\ Phys.\ Suppl.\ 168 (2007) 276.

\bibitem{datta-gupta} M.\ Panero, \PRL 103 (2009) 23001;\\
S.\ Datta and S.\ Gupta, \PR D 82 (2010) 114505. 

\bibitem{Kapusta} See e.g., J.\ I.\ Kapusta, {\sl Finite-Temperature Field
Theory}, Cambridge University Press, Cambridge 1989. 

\bibitem{Linde} A.\ D.\ Linde, \PL B 96 (1880) 289;\\
D.\ J.\ Gross, R.\ D.\ Pisarski and L.\ G.\ Yaffe, Rev.\ Mod.\ Phys.\ 53
(1981) 43. 

\bibitem{Arnold94} P.\ Arnold and C.-X.\ Zhai, \PR D50 (1994) 7609;\\
B.\ Kastening and C.-X.\ Zhai, \PR D52 (1995) 7232.

\bibitem{Laine} K.\ Kajantie et al., \PR D 67 (2003) 105008.

\bibitem{Patkos} F.\ Karsch, A.\ Patkos and P.\ Petreczky, \PL B 401 (1997) 69.

\bibitem{HTL} J.\ O.\ Andersen, M.\ Strickland and N.\ Su, 
JHEP 1008 (2010) 113.

\bibitem{resum} For reviews, see J.-P.\ Blaizot, \NP A 702 (2002) 99c;\\
A.\ Rebhan,  \NP A 702 (2002) 111c;\\
A.\ Peshier, \NP A 702 (2002) 128c.

\bibitem{L-Z} K.\ Lichtenegger and D.\ Zwanziger, \PR D 78 (2008) 034038.

\bibitem{Strick2} J.\ O.\ Andersen et al., \PL B 696 (2011) 468

\bibitem{bag} A.~Chodos et al.\,Phys.~Rev. D9 (1974) 3471.

\bibitem{Baacke} J.\ Baacke, Act.\ Phys.\ Polon.\ B 8 (1977) 625.

\bibitem{AH} M.\ Asakawa and T.\ Hatsuda, \NP A 610 (1996) 470c.

\bibitem{Leut} H.\ Leutwyler, in {\sl QCD - 20 Years Later},
P.\ M.\ Zerwas and H.\ Kastrup (Eds.), World Scientific, Singapore 1993,
pp.\ 693.

\bibitem{SVZ}  M.\ A.\ Shifman, A.\ I.\ Vainshtein and V.\ I.\ Zakharov,
\NP B 147 (1979) 385.

\bibitem{DEM} D.\ E.\ Miller, Phys.\ Rept.\ 443 (2007) 55.

\bibitem{Zwanziger} D.\ Zwanziger, \PRL 94 (2005) 182301.

\bibitem{Rob} R.\ Pisarski, \PR  D 74 (2006) 121703(R);\\
O.\ Andreev, \PR D 76 (2007) 087702.

\bibitem{Megias} E.\ Megias, E.\ Ruiz Arriola and L.\ L.\ Salcedo,
\PR D 80 (2009) 056005.

\bibitem{Golo93} V.\ Goloviznin and H.\ Satz, \ZP C57 (1993) 671.

\bibitem{Peshier} A.\ Peshier et al., \PR D 54 (1996) 2399.

\bibitem{Brau} F.\ Brau and F.\ Buisseret, \PR D 79 (2009) 114007. 

\bibitem{LH} P.\ L\'evai and U.\ Heinz, \PR C 57 (1998) 1879.

\bibitem{Gia} For recent work, see e.g., F.\ Giacosa,
arXiv:1009.4588 [hep-ph] 2010.

\bibitem{Mathieu} V.\ Mathieu, PoS QCD-TNT09:024 (2009).

\bibitem{Yang} C.-N.\ Yang, \PR 77 (1950) 242.

\bibitem{CCMS} P.\ Castorina et al., \EP 66 (2010) 207.

\end{thebibliography}
\end{document}